\documentstyle[seceq,preprint,epsbox]{jpsj}

\def\runtitle{Effect of Umklapp Scattering on Magnetic Field Penetration Depth 
in High-$T_{\rm c}$ Cuprates}
\def\runauthor{Takanobu {\sc Jujo}}

\title
{Effect of Umklapp Scattering on Magnetic Field Penetration Depth 
in High-$T_{\rm c}$ Cuprates}

\author
{Takanobu {\sc Jujo}\footnote{E-mail: jujo@ton.scphys.kyoto-u.ac.jp}}

\inst
{Department of Physics, Kyoto University, Kyoto 606-8502}

\recdate{May 14, 2001}

\abst
{The renormalization of the magnetic field penetration depth $\lambda$ 
owing to the electron-electron correlation is discussed with 
its application to high-$T_{\rm c}$ cuprates. 
The formula for the current carried by quasiparticle 
with the Umklapp scattering is derived, on the basis of which 
we investigate how the value of $\lambda^{-2}$ deviates from 
that of $n/m^*$ where $n$ and $m^*$ are the carrier density 
and the effective mass respectively. 
Although this deviation is small in the case 
of weak momentum dependence of the vertex, this 
is large and negative owing to the non-negligible 
value of the backflow in the case of the strong 
antiferromagnetic spin fluctuation. 
The observed doping dependence of $\lambda^{-2}$ in 
high-$T_{\rm c}$ cuprates, specifically a peak structure 
at the slightly overdoped region, is explained by the 
analytical consideration and the numerical calculation 
based on the perturbation theory and the spin fluctuation theory. 
The consistency between $\lambda^{-2}$ and 
${\rm d}\lambda^{-2}/{\rm d}T$ at absolute zero, which 
is the problem the isotropic model fails to explain, 
is also obtained by our theory. 
}

\kword
{Fermi liquid theory, high-$T_{\rm c}$ cuprates, 
current of quasiparticles, Umklapp scattering, 
electron-electron correlation, 
vertex correction, magnetic field penetration depth, 
perturbation theory, antiferromagnetic spin fluctuation
}

\begin{document}
\sloppy
\maketitle

\section{Introduction}

In high-$T_{\rm c}$ cuprates some transport phenomena 
show characteristic behaviors, especially in 
the optimal and the underdoped regions which 
do not seem to be explained by the 
conventional Fermi liquid theory. 
For examples, the resistivity is proportional to the temperature $T$, 
rather than $T^2$.~\cite{rf:1}
The Hall coefficient is temperature dependent and increases 
as the doping level decreases.~\cite{rf:2}
It is also seen that not only these transport phenomena, 
but also the magnetic properties like the nuclear spin-lattice relaxation 
rate and the one-particle spectrum show the pseudo gap 
phenomenon which is the gap like behavior with no long range order. 
The magnetic field penetration depth $\lambda$ in the high-$T_{\rm c}$ 
cuprates attracts attention because $\lambda$ is 
long in the underdoped region where the pseudogap phenomena appear 
and then some authors suggest the low carrier density based on the relation 
$\lambda^{-2} \propto n/m^*$ where $n$,$m^*$ are the carrier density and 
the effective mass, respectively,~\cite{rf:3} and a connection 
between long $\lambda$ and the pseudogap phenomena.~\cite{rf:4} 

In the above phenomena it is found in the early stage 
of the study on the high-$T_{\rm c}$ cuprates that the 
$T$-linear resistivity is explained by the 
fact that the lifetime of the quasiparticles is proportional to $T$ 
in the presence of the strong antiferromagnetic spin 
fluctuation.~\cite{rf:5,rf:6} 
This means that it is sufficient to consider an one-particle property like 
the imaginary part of the self-energy. 
The pseudo gap phenomena can also be explained by the 
superconducting fluctuation with the two dimensional and 
the strong coupling nature of the system.~\cite{rf:7,rf:8} 
In this case the absolute value of the imaginary part of 
the self-energy shows the maximum at the Fermi level 
and therefore one particle spectrum decreases. 
On the other hand it is found that the interaction 
between the quasiparticles is essential for the 
explanation of the Hall coefficient.~\cite{rf:9} 
This work clarifies the importance of the interaction between 
the quasiparticles besides one-particle properties like the self-energy. 

As for the magnetic field penetration depth in high-$T_{\rm c}$ cuprates, 
there has been no satisfactory theories until now. 
As noted above it seems that $\lambda^{-2}$ is proportional 
to the doping concentration $\delta$, and from the combination 
with $\lambda^{-2} \propto n/m^*$ it seems to be reasonable 
to expect that the effective carrier density is identified 
with the doping concentration (i.e. $n \propto \delta$).~\cite{rf:3} 
The theories based on the $t$-$J$ model naturally accept this conjecture 
because of no double occupation. 
There is no need to explain why $\lambda^{-2}$ is so small 
but it needs only to take it as an external parameter 
in these theories.~\cite{rf:10} 
However the doping dependences of $\lambda^{-2}$ show that 
$\lambda^{-2}$ increases with $\delta$ in the underdoped and 
optimal regions, and it has a peak at the slightly 
overdoped region, and then it decreases as $\delta$ increases 
in the overdoped region.~\cite{rf:11} 
It is obvious that the latter behavior cannot be explained by 
the theory based on the $t$-$J$ model. 
On the other hand $n$ should be proportional to $1-\delta$ 
in the Fermi gas theory because $\delta=0$ means the half-filled, 
and therefore based on $\lambda^{-2} \propto n/m^*$ 
it can explain the overdoped region but cannot explain the optimal 
and underdoped regions. 
If we take account of the interaction between the quasiparticles, 
it is known that $\lambda^{-2} \propto n(1+F^s_1/3)/m^*$.~\cite{rf:12} 
Here $F^s_1$ is one of the Landau parameters and this expression 
is based on the isotropic model. 
Though the behavior of this parameter is not known, 
it is seemingly possible to explain the doping dependence of 
$\lambda^{-2}$ by controlling the parameter $F^s_1$. 
However it is shown that this parameter theory fails 
if we try to maintain the consistency between the values of 
$\lambda^{-2}$ and ${\rm d}\lambda^{-2}/{\rm d}T$
at absolute zero.~\cite{rf:13}
These considerations indicate that the isotropic 
formalism presented by Leggett should be reexamined 
to investigate the realistic metallic system like the 
high-$T_{\rm c}$ cuprates, and it cannot be permitted 
to take $F^s_1$ as an external parameter either, necessitating 
the investigation of the quantities corresponding to $F^s_1$ in detail. 

The Umklapp scattering, a main interest in this paper, 
is important for the transport phenomena. 
In the previous paper a systematic discussion on 
how the many body effect appears in the superconducting state, 
is presented.~\cite{rf:14} 
Some examples presented there, are 
the magnetic field penetration depth 
and the optical conductivity. 
At absolute zero the expression of the magnetic field penetration 
depth contains the interaction between 
quasiparticles in the same way as in the cyclotron resonance 
frequency~\cite{rf:15} and the Drude weight.~\cite{rf:16,rf:17} 
In ref.18 it is shown that the absence of the Umklapp 
scattering leads to the infinite conductivity. 
Although this fact can be shown by using the Ward-Takahashi 
identity related with the momentum conservation, 
the essential point in ref.18 is that the formula 
for the resistivity is written explicitly by the Umklapp process. 
The formula for the current which is explicitly 
written by the Umklapp scattering 
has not been known. The derivation of this formula is given 
in this paper. 
With regard to high-$T_{\rm c}$ cuprates it is needed to 
clarify the relation between the Umklapp scattering 
and the antiferromagnetic spin fluctuation and 
it is shown that the strong antiferromagnetic spin 
fluctuation leads to the strong Umklapp scattering 
on the current. 

In \S2 the formula for the current carried by 
quasiparticles written by the Umklapp scattering 
is derived and the other property of the current is 
briefly discussed. 
In \S3 the model and the approximations for the 
analytical and the numerical calculations are presented. 
Some discussions on the influence of the 
superconducting transition is also given and 
the validity of the calculation of the current 
in the normal state is shown. 
In \S4 the analytical investigation for the current is 
presented in the two specific cases. 
One is the case where the momentum dependence of the 
irreducible four point vertex is weak and 
the other is the case with the strong antiferromagnetic 
spin fluctuation. 
It is shown that the discussions based on $\lambda^{-2} \propto 
n/m^*$ is approximately valid only in the former case. 
In the latter case the behaviors of the current 
differ with the others depending on the points 
on the Fermi surface and it is shown that the 
current on the points with the strong antiferromagnetic 
spin fluctuation is decreased much. 
How the effect of the Umklapp scattering on the current 
depends on the irreducible four point vertex is also presented 
and it is shown that it is largest in the case of the strong 
antiferromagnetic spin fluctuation and is smallest 
in the case of the ferromagnetic spin fluctuation. 
It is also presented that the backflow in the case of the strong 
antiferromagnetic spin fluctuation has an opposite 
sign to the bare velocity. 
In \S5 the numerical calculations with the self-consistent 
second order perturbation theory (SC-SOPT) and the 
fluctuation exchange approximation (FLEX) are presented. 
The results which are consistent with the experiments 
in the respects of the doping dependence of $\lambda^{-2}$ 
at absolute zero and the consistency between $\lambda^{-2}$ 
and ${\rm d}\lambda^{-2}/{\rm d}T$, are obtained. 
In \S6 a brief consideration on the relation between the 
superconducting fluctuation and $\lambda^{-2}$ is given 
and it is indicated that the small $\lambda^{-2}$ at absolute 
zero does not directly mean the large thermal fluctuation. 
In \S7 the summary and the discussion are given. 
   
In this paper we put $\hbar=k_{\rm B}=c=1$.

\section{Current Carried by Quasiparticles}

The current carried by quasiparticles 
plays an important role in physical 
quantities of the collisionless region in the normal state, 
for example, the cyclotron resonance frequency 
and the Drude weight. 
This is usually written by using 
only quantities defined on the Fermi surface as 
\begin{equation}
j^*_{\mib k\mu}=v^*_{\mib k\mu}+\frac{1}{V}\sum_{\mib k}
f_{\mib k,\mib k'}\delta(\xi_{\mib k'}^*)v^*_{\mib k'\mu}.
\end{equation}
Here $\mu$ is the index of spatial dimensions, $V$ is the volume 
of the system, $\xi^*_{\mib k}$ is the dispersion of the quasiparticles 
including the chemical potential, $v^*_{\mib k\mu}$ 
is the renormalized velocity of quasiparticles 
(${\mib v}^*_{\mib k}=\frac{\partial \xi^*_{\mib k}}{\partial \mib k}$) 
and $f_{\mib k,\mib k'}=z_{\mib k}
\Gamma^{\omega}_{\mib k,\mib k'}z_{\mib k'}$ is the interaction 
between quasiparticles ($\Gamma^{\omega}$ is the 
$\omega$-limit of reducible four-point vertex and 
the notation about $\omega$-limit and $k$-limit is the same as 
in ref.19). 
In the superconducting state the magnetic field penetration 
depth is also written by using this quantity at absolute zero. 
As derived in the previous paper,~\cite{rf:14} 
the magnetic field penetration depth $\lambda$ 
at finite temperature is written as 
\begin{equation}
\left(\frac{1}{\lambda^2}\right)_{\mu\nu} \propto
\int_{\rm FS}\frac{{\rm d}S_k}{4\pi^3|{\mib v}^*(\mib k)|}
v^*_{\mu}(\mib k)(1-Y(\mib k;T))\bar{v}^*_{\nu}(\mib k;T). 
\end{equation}
Here $\int_{\rm FS}{\rm d}S_k$ is the integral 
over the Fermi surface and 
$Y(\mib k;T)$ is Yosida function 
\begin{equation}
Y(\mib k;T)=\int{\rm d}\xi^*_{\mib k}\left(
-\frac{\partial f(E^*_{\mib k})}{\partial E^*_{\mib k}}\right), 
\end{equation}
$f(x):=1/({\rm e}^{x/T}+1)$, 
$E^*_{\mib k}=\sqrt(\xi^{*2}_{\mib k}+\Delta^2_{\mib k})$, $\Delta_{\mib k}$ 
is the superconducting gap and 
$\bar{v}^*_{\nu}(\mib k;T)$ satisfies 
the following integral equation, 
\begin{equation} 
\bar{v}^*_{\nu}(\mib k;T)=j^*_{\nu}(\mib k)
-\int_{\rm FS}\frac{{\rm d}S_{\mib k'}}{4\pi^3|{\mib v}^*(\mib k')|}
f_{\mib k,\mib k'}Y(\mib k';T)\bar{v}^*_{\nu}(\mib k';T), 
\end{equation}
or
\begin{equation} 
\bar{v}^*_{\nu}(\mib k;T)=v^*_{\nu}(\mib k)
+\int_{\rm FS}\frac{{\rm d}S_{\mib k'}}{4\pi^3|{\mib v}^*(\mib k')|}
A_{\mib k,\mib k'}(1-Y(\mib k';T))\bar{v}^*_{\nu}(\mib k';T). 
\end{equation}
Then it can be seen that $\bar{v}^*_{\nu}(\mib k;T)=
j^*_{\mib k\nu}$ at absolute zero because of $Y(\mib k;T=0)=0$. 
Here $A_{\mib k,\mib k'}=z_{\mib k}\Gamma^k_{\mib k,\mib k'}z_{\mib k'}$ 
($\Gamma^k_{\mib k,\mib k'}$ is the $k$-limit of the reducible 
four point vertex) and this satisfies the following equation: 
\begin{equation}
f_{\mib k,\mib k'}=A_{\mib k,\mib k'}+\frac{1}{V}\sum_{\mib k''}
A_{\mib k,\mib k'}\left(-\frac{\partial f(\xi^*_{\mib k''})}
{\partial \xi^*_{\mib k''}}\right)f_{\mib k'',\mib k'}. 
\end{equation}

Here we briefly explain why $\lambda^{-2}$ is not generally 
written as $n/m^*$, and how the many body effect enters. 
The electromagnetic response kernel is written as 
\begin{equation}
K_{\mu\nu}=-\int_{k}v_{\mib k \mu}
(GG+FF)_{\mib k}(\epsilon)\Lambda_{\mib k \nu}(\epsilon)
-\left(\frac{n}{m}\right)_{\mu\nu}. 
\end{equation}
Here $G$ and $F$ are the normal and the anomalous Green's function, 
respectively, in the superconducting state in the usual sense,$^{19)}$, 
$m$ is the bare mass, $v_{\mib k\mu}$ is the bare velocity 
and $\Lambda_{\mib k\nu}(\epsilon)$ is the three point vertex 
which satisfies the integral equation with the interaction included. 
$\int_k := \int\frac{{\rm d}\epsilon}{4\pi{\rm i}V}\sum_{\mib k}$ 
is used in the zero temperature formalism hereafter. 
In the case of the finite temperature formalism 
$\int_k := \int\frac{{\rm d}\epsilon}{2\pi V}\sum_{\mib k}$ is 
used. 
The first term of the r.h.s. of eq.(2.7) is often called 
paramagnetic term and the second term is called 
diamagnetic term. 
In the usual textbook~\cite{rf:20} it is explained that the 
paramagnetic term vanishes at absolute zero in the superconducting
state. 
This is valid in the case where the superconducting gap grows 
in the whole Fermi sea and then the integral which includes 
$GG+FF$ reduced to Yosida function and vanishes at absolute 
zero. 
However in all superconductors the superconducting gap 
grows only near the Fermi surface and the incoherent part 
contributes to the integral which does not vanish in 
the lattice system. 
Then the eq.(2.7) is reduced to the following equation 
at absolute zero, instead of to $-n/m$. 
\begin{equation}
K_{\mu\nu}=-\int_k
v_{\mib k \mu}(GG)^{\rm inc}_{\mib k}(\epsilon)
\Lambda_{\mib k \nu}(\epsilon)
-\left(\frac{n}{m}\right)_{\mu\nu}. 
\end{equation}
In the same way $\Lambda_{\mib k\nu}(\epsilon)$ 
satisfies the following integral equation. 
\begin{equation}
\Lambda_{\mib k \nu}(\epsilon)=v_{\mib k \nu}
+\int_{k'}I_{\mib k,\mib k'}(\epsilon,\epsilon')
(GG)^{\rm inc}_{\mib k'}(\epsilon')\Lambda_{\mib k' \nu}(\epsilon'). 
\end{equation}
($I_{\mib k,\mib k'}(\epsilon,\epsilon')$ is the irreducible 
four-point vertex.) 
Then from eqs.(2.7) and (2.9) 
the formalism of magnetic field penetration depth 
based on Fermi liquid theory is derived by 
integrating the high energy part to derive the low energy 
dynamics as investigated in detail in ref.14. 
In the specific case where the momentum is conserved 
it can be shown that the paramagnetic term exactly 
vanishes at absolute zero as follows. 
In this case Ward-Takahashi identity which reflect 
the momentum conservation is written as 
\begin{equation}
\Lambda_{\mib k \nu}(\epsilon)=
\left(1-\frac{\partial \Sigma^n_{\mib k}(\epsilon)}
{\partial \epsilon}\right)v_{\mib k \nu}. 
\end{equation}
($\Sigma^n_{\mib k}(\epsilon)$ is the normal self-energy.) 
This relation holds at arbitrary value of $\epsilon$. 
Then the paramagnetic term is 
\begin{eqnarray}
-\int_k v_{\mib k \mu}(GG)^{\rm inc}_{\mib k}(\epsilon)
\left(1-\frac{\partial \Sigma^n_{\mib k}(\epsilon)}
{\partial \epsilon}\right)v_{\mib k \nu}
&=&\int_{k}v_{\mib k \mu}v_{\mib k \nu}
\frac{\partial G_{\mib k}(\epsilon)}{\partial \epsilon} \\
&=& 0. 
\end{eqnarray}
This is the direct proof of the disappearance of the 
paramagnetic term in the case where the momentum conservation 
holds, which is not via the Fermi liquid parameter like $F^s_1$. 

Next we discuss the role of the Umklapp scattering. 
Hereafter in this section the discussion is given by using the 
Green's function in the normal state. 
In the superconducting state equations become more 
complicated, but the main results are not altered 
as partly discussed in ref.14 about the renormalization 
factor and the renormalized velocity and in \S3.1. 
By using the fact that the renormalized velocity 
and the renormalization factor is written as 
($k$ is a simplified notation of $(\mib k,\epsilon)$ 
in the zero temperature formalism), 
\begin{equation}
v^*_{\mib k\mu}=z_{\mib k}(v_{\mib k\mu}
+\int_{k'}\Gamma^k_{\mib k,\mib k'}(0,\epsilon')
[G(k')^2]^k v_{\mib k'\mu}) 
\end{equation}
and
\begin{equation}
z^{-1}_{\mib k}=1+\int_{k'}
\Gamma^{\omega}_{\mib k,\mib k'}(0,\epsilon')
[G(k')^2]^{\omega}
\end{equation}
respectively, and also noting that 
\begin{equation}  
[G(k)^2]^{\omega}-[G(k)^2]^k=2\pi {\rm i}\delta(\epsilon)
\delta(\xi^*_{\mib k}), 
\end{equation}
eq.(2.1) is written as 
\begin{equation}
j^*_{\mib k \mu}
=v_{\mib k \mu}+z_{\mib k}\int_{k'}
\Gamma^{\omega}_{\mib k,\mib k'}(0,\epsilon')
[G(k')^2]^{\omega}(v_{\mib k' \mu}-v_{\mib k \mu}),
\end{equation}
or 
\begin{equation}
j^*_{\mib k \mu}=v_{\mib k \mu}+z_{\mib k}\left(
\frac{{\rm d} \Sigma(k)}{{\rm d} \mib k}+
\frac{\partial \Sigma(k)}{\partial \epsilon}v_{\mib k \mu}
\right)_{\epsilon =0}. 
\end{equation}
Here $\frac{{\rm d} \Sigma(k)}{{\rm d} \mib k}$ is the 
same notation as in ref.21 and it means the 
derivative with the Fermi surface also transformed, 
in the contrary to $\frac{\partial \Sigma(k)}{\partial \mib k}$, 
the derivative with the Fermi surface fixed. 
Reflecting the simultaneous transformation of the Fermi surface, 
$\frac{{\rm d} \Sigma(k)}{{\rm d} \mib k}$ corresponds to 
the $\omega$-limit, and $\frac{\partial \Sigma(k)}{\partial \mib k}$ 
to the $k$-limit owing to the inclusion of the 
discontinuity on the Fermi surface. 

The quantities, $\frac{{\rm d} \Sigma(k)}{{\rm d} \mib k}$ and 
$\frac{\partial \Sigma(k)}{\partial \epsilon}$, satisfy 
the following integral equations, respectively. 
\begin{equation}
\frac{{\rm d} \Sigma(k)}{{\rm d} \mib k}=
\int_{k'}I(k,k')[G(k')^2]^{\omega}\left({\mib v}_{\mib k'}+
\frac{{\rm d} \Sigma(k')}{{\rm d} \mib k'}\right)
\end{equation}
and
\begin{equation}
\frac{\partial \Sigma(k)}{\partial \epsilon}=
-\int_{k'}I(k,k')[G(k')^2]^{\omega}\left(1-
\frac{\partial \Sigma(k')}{\partial \epsilon'}\right). 
\end{equation}
Then it is verified that ${\mib w}_{\mib k}(\epsilon):=
\frac{{\rm d} \Sigma(k)}{{\rm d} \mib k}+
\frac{\partial \Sigma(k)}{\partial \epsilon}{\mib v}_{\mib k}$ 
satisfies the following integral equation. 
\begin{eqnarray}
{\mib w}_{\mib k}(\epsilon)
&=&
{\mib u}_{\mib k}(\epsilon)+\int_{k'}I(k,k')[G(k')^2]^{\omega}
{\mib w}_{\mib k'}(\epsilon') \\
&=&
{\mib u}_{\mib k}(\epsilon)+\int_{k'}\Gamma^{\omega}(k,k')
[G(k')^2]^{\omega}{\mib u}_{\mib k'}(\epsilon'). 
\end{eqnarray}
Here 
\begin{equation}
u_{\mib k \mu}(\epsilon)=\int_{k'}I(k,k')[G(k')^2]^{\omega}
\left(1-\frac{\partial \Sigma(k')}{\partial \epsilon'}\right)
(v_{\mib k' \mu}-v_{\mib k \mu}). 
\end{equation}
This is rewritten as 
\begin{eqnarray}
{\mib u}_{\mib k}(\epsilon) & = & 
\int_{k'}I(k,k')\left(-\frac{\partial G(k')}{\partial \epsilon'}\right)
{\mib v}_{\mib k'}+\frac{\partial \Sigma (k)}
{\partial \epsilon}{\mib v}_{\mib k} \\
& = & \int_{k'}\int_{q}|\Gamma(k,k';q)|^2
G(k-q)G(k'+q)\left(-
\frac{\partial G(k')}{\partial \epsilon'}\right)
({\mib v}_{\mib k'+\mib q}+{\mib v}_{\mib k-\mib q}
-{\mib v}_{\mib k'}-{\mib v}_{\mib k}). 
\end{eqnarray}
This equation shows that ${\mib u}_{\mib k}(\epsilon)=\mib 0$ 
in the absence of the Umklapp process and the concrete 
examples are given in \S3 in the case 
of SC-SOPT and FLEX approximations. 
By using the above quantities the backflow term is written by 
the difference between the two kinds of 
the momentum derivative of self-energy. 
\begin{equation}
\frac{{\rm d} \Sigma(k)}{{\rm d} \mib k}-
\frac{\partial \Sigma(k)}{\partial \mib k}
=\frac{\mib B_{\mib k}}{z_{\mib k}}. 
\end{equation}
Here ${\mib B}_{\mib k}$ is the backflow term. 

In ref.15 $j^*_{\mib k\mu}$ is calculated based on  
second order perturbation theory with respect to the on-site Coulomb 
interaction. It is mentioned there that the 
four-point interaction vertex is taken to satisfy the 
conservation law and they obtained the results that 
the vertex correction make the positive value in some parameter regions. 
If this holds, it follows that the value of current carried by quasiparticles 
exceeds the value of the bare electron's velocity. 
However it is not considered to be possible for the interacting electrons 
to carry the current which exceeds the non-interacting case 
because if it were possible, gathering the momentum 
from the crystalline lattice and carrying it would occur. 
If we see the general expressions above it is obvious 
that their treatment is incorrect. 
To satisfy the conservation law and get the correct value of 
$j^*_{\mib k\mu}$, it is needed to use eq.(2.24) as the vertex correction 
with $|\Gamma(k,k';q)|^2$ replaced by $U^2$ in the case of the second order 
perturbation. 
The fact that $j^*_{\mib k\mu}$ does not exceeds 
$v_{\mib k\mu}$ is made clear by eq.(2.16) 
with the notion that $v_{\mib k\mu}$ is an odd function 
with respect to $\mib k$ and $\int_{k'}
\Gamma^{\omega}_{\mib k,\mib k'}(0,\epsilon')[G(k')^2]^{\omega} > 0$ 
in the Fermi liquid system. 
Then $j^*_{\mib k\mu}<v_{\mib k\mu}$ in $\mib k$ with 
$v_{\mib k \mu}>0$, and this is the basic inequality 
which can be obtained from the above general discussion. 
This inequality also guarantee that $\lambda^{-2}$ is always 
smaller than $4\pi e^2 n/m$. 

By taking the above results into account 
we consider how the currents $j^*$ in the lattice system 
changes from the value $j^*=v$ which is universal in the case of 
the continuum. 
If we consider this correction in terms of the renormalization factor 
($z$), it can be seen that this quantity is 
changes as $(1-z^{-1})/z^{-1}$ plus the extra contribution 
proportional to $z$ which depends on the system considered. 
The presence of this extra contribution 
violates the relation $\lambda \propto n/m^*$ and 
expresses one of the characteristics of the system like the 
momentum dependence of the interaction between the quasiparticles. 
This property in high-$T_{\rm c}$ cuprates is investigated in detail 
in \S4 and 5. 

\section{Model and Approximation}

We take the following Hubbard Hamiltonian with 
on-site Coulomb interaction. 
\begin{equation}
{\cal H}=\sum_{\mib k,\sigma} \xi_{\mib k} 
c^{\dagger}_{\mib k \sigma}c_{\mib k \sigma}
+U\sum_{\mib k,\mib k',\mib q}
c^{\dagger}_{\mib k+\mib q \uparrow}c^{\dagger}_{\mib k'-\mib q \downarrow}
c_{\mib k' \downarrow}c_{\mib k \uparrow}. 
\end{equation}
Here $\xi_{\mib k}$ is the dispersion of electrons 
which is taken similar to that in ref.9. 
\begin{equation}
\xi_{\mib k}=-2t({\rm cos}k_x+{\rm cos}k_y)+4t'{\rm cos}k_x{\rm cos}k_y
-2t''({\rm cos}2k_x+{\rm cos}2k_y). 
\end{equation}
In the numerical calculation we put $t'/t=0.16$, $t''/t=0.20$, 
$U/t=7.0$ and $t=1$. 

The self-energy within SC-SOPT is 
\begin{equation}
\Sigma(k)=\frac{T}{V}\sum_{q}\chi(q)G(k-q). 
\end{equation}
Here $k=(\mib k,{\rm i}\epsilon_n)$ ($\epsilon_n$ 
is the Matsubara frequency), 
\begin{equation}
G(k)=\frac{1}{{\rm i}\epsilon_n-\xi_{\mib k}-\Sigma(k)}
\end{equation}
and 
\begin{equation}
\chi(q)=-\frac{T}{V}\sum_{k'}G(k')G(k'+q). 
\end{equation}
The irreducible four point vertex is 
\begin{eqnarray}
I(k,k')&=&
\frac{\delta \Sigma(k)}{\delta G(k')} \nonumber\\
&=&2\chi(k-k')+\phi(k+k'). 
\end{eqnarray}
The first line of this equation follows the conserving 
approximation by Baym and Kadanoff.~\cite{rf:22}. 
Here
\begin{equation}
\phi(q)=-\frac{T}{V}\sum_{k'}G(k')G(q-k'). 
\end{equation}
$\Gamma^{(a)}_{\mib k,\mib k'}(\epsilon,\epsilon')$ and 
$\Gamma^{(b)}_{\mib k,\mib k'}(\epsilon,\epsilon')$
in the basic formalism in Appendix are given by the following 
expressions. 
\begin{equation}
\Gamma^{(a)}_{\mib k,\mib k'}(\epsilon,\epsilon')
=2\chi^R_{\mib k-\mib k'}(\epsilon-\epsilon')
\end{equation}
and
\begin{equation}
\Gamma^{(b)}_{\mib k,\mib k'}(\epsilon,\epsilon')
=\phi^R_{\mib k+\mib k'}(\epsilon+\epsilon'). 
\end{equation}
By using the above $I(k,k')$, ${\mib u}_{\mib k}(\epsilon)$ 
which expresses the Umklapp term is explicitly written as 
\begin{equation}
{\mib u}_{\mib k}(\epsilon)=
U^2\int_{k'}\int_{q}G_{\mib k-\mib q}(\epsilon-\omega) 
G_{\mib k'+\mib q}(\epsilon'+\omega)
\frac{\partial G_{\mib k'}(\epsilon')}{\partial \epsilon'}
(\mib v_{\mib k}+\mib v_{\mib k'}-\mib v_{\mib k'+\mib q}
-\mib v_{\mib k-\mib q})
\end{equation}
As can be easily seen by explicit calculation 
$-{\mib v}_{\mib k-\mib q}$ term 
and $-{\mib v}_{\mib k'+\mib q}$ in the above equation 
originate from $2\chi(q)$ term in $I(k,k')$, 
${\mib v}_{\mib k'}$ term from $\phi(q)$ and 
${\mib v}_{\mib k}$ term from $\partial \Sigma(k)/\partial \epsilon$. 

The basic equations in the FLEX approximation are 
(for an example, see ref.23), 
\begin{equation}
\Sigma(k)=\frac{T}{V}\sum_{q}V(q)G(k-q),
\end{equation}
\begin{equation}
V(q)=\frac{3}{2}U^2\frac{\chi(q)}{1-U\chi(q)}+
\frac{1}{2}U^2\frac{\chi(q)}{1+U\chi(q)}-U^2\chi(q). 
\end{equation}
The irreducible four-point vertex in this approximation 
is 
\begin{eqnarray}
I(k,k')
&=&\frac{\delta \Sigma(k)}{\delta G(k')} \nonumber\\
&=&V(k-k')-\frac{T}{V}\sum_{q}W(q)[G(k'-q)G(k-q)+
G(k'+q)G(k-q)]. 
\end{eqnarray}
Here 
\begin{equation}
W(q)=\frac{3}{2}U^2\frac{1}{|1-U\chi(q)|^2}
+\frac{1}{2}U^2\frac{1}{|1+U\chi(q)|^2}-U^2. 
\end{equation}
$\Gamma^{(a)}_{\mib k,\mib k'}(\epsilon,\epsilon')$ and 
$\Gamma^{(b)}_{\mib k,\mib k'}(\epsilon,\epsilon')$
in the basic formalism in Appendix are given by the following 
expressions. 
\begin{eqnarray}
\Gamma^{(a)}_{\mib k,\mib k'}(\epsilon,\epsilon')&=&
V^R_{\mib k -\mib k'}(\epsilon-\epsilon')-
\frac{1}{V}\sum_{\mib q}\int\frac{{\rm d}\omega}{2\pi}
W_{\mib q}(\omega)[
{\rm tanh}\frac{\epsilon-\omega}{2T}
{\rm Im}G^R_{\mib k - \mib q}(\epsilon -\omega)
G^A_{\mib k' -\mib q}(\epsilon'-\omega) \nonumber\\
&+&{\rm tanh}\frac{\epsilon'-\omega}{2T}
G^R_{\mib k - \mib q}(\epsilon -\omega)
{\rm Im}G^R_{\mib k' -\mib q}(\epsilon'-\omega)] 
\end{eqnarray}
and 
\begin{eqnarray}
\Gamma^{(b)}_{\mib k,\mib k'}(\epsilon,\epsilon')&=&
-\frac{1}{V}\sum_{\mib q}\int\frac{{\rm d}\omega}{2\pi}
W_{\mib q}(\omega)[
{\rm tanh}\frac{\epsilon-\omega}{2T}
{\rm Im}G^R_{\mib k-\mib q}(\epsilon-\omega)
G^R_{\mib k'+\mib q}(\epsilon'+\omega) \nonumber\\
&+&{\rm tanh}\frac{\epsilon'+\omega}{2T}
G^R_{\mib k-\mib q}(\epsilon-\omega)
{\rm Im}G^R_{\mib k'+\mib q}(\epsilon'+\omega)]. 
\end{eqnarray}

By using the above $I(k,k')$, ${\mib u}_{\mib k}(\epsilon)$ 
which expresses the Umklapp term is explicitly written as 
\begin{equation}
{\mib u}_{\mib k}(\epsilon)=
\int_{k'}\int_{q}W_{\mib q}(\omega)
G_{\mib k-\mib q}(\epsilon-\omega) 
G_{\mib k'+\mib q}(\epsilon'+\omega)
\frac{\partial G_{\mib k'}(\epsilon')}{\partial \epsilon'}
(\mib v_{\mib k}+\mib v_{\mib k'}-\mib v_{\mib k'+\mib q}
-\mib v_{\mib k-\mib q})
\end{equation}
As can be easily seen by explicit calculation 
$-{\mib v}_{\mib k-\mib q}$ term in the above equation 
originate from $V(q)$ term in $I(k,k')$, 
$-{\mib v}_{\mib k'+\mib q}$ term from 
the second term of r.h.s. of eq.(3.13), 
${\mib v}_{\mib k'}$ term from 
the third term of r.h.s. of eq.(3.13) and 
${\mib v}_{\mib k}$ term from $\partial \Sigma(k)/\partial \epsilon$. 

By using the above approximations, 
it is shown that the momentum dependence 
of $V^R_{\mib q}(0)$ is remarkable at $\mib q=(\pi,\pi)$. 
On the other hand, although $\chi^R_{\mib q}(0)$ has a 
small peak around $\mib q=(\pi,\pi)$ and $\phi^R_{\mib q}(0)$ 
at $\mib q=(0,0)$, it can be seen that 
the momentum dependence of the 
irreducible four point vertex in SC-SOPT is rather weak 
as compared with that in FLEX. 

\subsection{Influence of Superconducting Transition}
Here we briefly discuss how the electronic properties 
differ between the normal state and the superconducting state, 
and take the renormalization factor and the damping rate 
of the quasiparticles as the representative quantities 
because the renormalization factor is related to 
the vertex corrections to the current 
by the real part of the irreducible four point vertex and 
the damping rate of the quasiparticles is related to 
the vertex correction in the hydrodynamic region 
by the imaginary part. 
It is known that the imaginary part of the vertex 
caused by the electron-electron interaction becomes small 
at the low energy in the superconducting state because of the decrease 
of the scattering accompanied by the gap formation, while 
the real part is not so affected at low energy except for the uniform 
magnetic susceptibility in the case of the singlet pairing. 

\begin{figure}
\epsfile{file=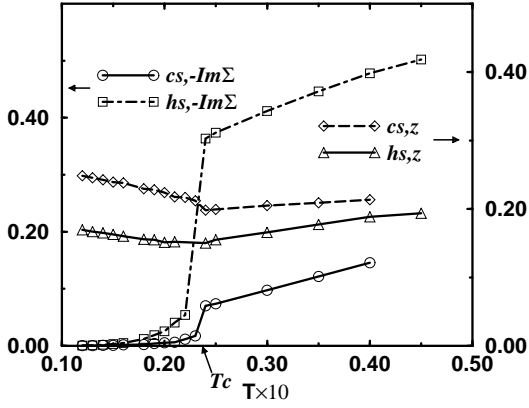,height=6.0cm}
\caption{$-{\rm Im}\Sigma^R_{\mib k}(0)$ 
and the renormalization factor ($z$) at the cold spot ($cs$) 
and the hot spot ($hs$). 
$\delta=0.10$ and $T_{\rm c}=0.024$.}
\label{fig:1}
\end{figure}

The imaginary part of the self-energy and the renormalization 
factor calculated numerically in the FLEX approximation 
is shown in Fig.~\ref{fig:1}. 
From this figure it is shown that $-{\rm Im}\Sigma^R_{\mib k}(0)$ 
rapidly decreases below $T_{\rm c}$ as 
expected from the above consideration. 
A notable point concerning the damping rate, 
which is not related to the present discussion, 
is that the damping rate at the cold spot decreases 
linearly with the temperature and that at the hot spot 
decreases as $\sqrt T$. The former point causes 
the $T$-linear resistivity and the latter point 
does not affect the conductivity 
because the point with the smallest 
damping rate mainly contributes to the conductivity. 
 
On the other hand the renormalization factor 
almost remains constant as expected also, 
while the change in the spectrum of the spin fluctuation 
makes the minimum of this quantity at $T=T_{\rm c}$. 
These considerations justify the calculation 
of the current in the normal state to estimate 
the magnetic field penetration depth because the 
vertex correction for the current is the type 
of the function analogous to the renormalization factor. 

\section{Analysis of Umklapp Term}

\subsection{The case where the momentum dependence of 
$I_{\mib k,\mib k'}$ is weak} 

This case occurs when the shape of the Fermi surface 
is not so peculiar and the perturbation scheme is valid. 
Then it is easy to derive the following relation 
by using the fact that the velocity is an odd function 
of $\mib k$. 
\begin{eqnarray}
{\mib w}_{\mib k}(\epsilon)
&\simeq& {\mib u}_{\mib k}(\epsilon) \\
&\simeq& \int_{k'}I_{\mib k,\mib k'}(\epsilon,\epsilon')
\frac{\partial G_{\mib k'}(\epsilon')}
{\partial \epsilon'}{\mib v}_{\mib k} \\
&=&\frac{\partial \Sigma_{\mib k}(\epsilon)}
{\partial \epsilon}{\mib v}_{\mib k}. 
\end{eqnarray}
Then the current is written as 
\begin{eqnarray}
j^*_{\mib k \mu}
&\simeq&v_{\mib k \mu}
+z_{\mib k}\frac{\partial \Sigma_{\mib k}(\epsilon)}
{\partial \epsilon}|_{\epsilon=0}v_{\mib k \mu} \\
&=&z_{\mib k} v_{\mib k \mu}
\end{eqnarray}
This equation shows that 
the Umklapp term reduces to $(z_{\mib k}-1){\mib v}_{\mib k}$. 
The current and $\lambda^{-2}$ 
is only reduced by $z_{\mib k}$ and then 
the latter quantity is interpreted by writing 
$n/m^*$ as used $\it a priori$ in the some 
papers,~\cite{rf:3} where $m^*$ is the thermal mass which is  
enhanced by $1/z$. 
However it is considered that in the case of high-$T_{\rm c}$ 
the validity of this perturbation 
scheme is restricted in the overdoped region. 
This assertion is understood by noting the behavior 
of the nuclear spin relaxation rate and the resistivity 
measurements. 
Therefore the use of the form $n/m^*$ in the underdoped cuprates, 
which is taken in some papers, is not warranted. 

\subsection{The case of strong antiferromagnetic spin fluctuation}

This is the case which is relevant to the optimal and 
underdoped cuprates. 
In this case $W_{\mib q}(\omega)$ has a sharp peak at 
$\mib q \simeq \mib Q$ and $\omega \simeq 0$ ($\mib Q=(\pi,\pi)$). 
The fact that $W_{\mib q}$ takes a large value at a specific 
momentum reduces the integral including $W_{\mib q}$ 
to the one that is the integral of the odd function $v_{\mib k}$, 
and then the term including $W_{\mib q}$ can be neglected, 
as confirmed numerically. 
Then eqs.(2.20) and (2.22) (the corresponding equations in 
Appendix) is reduced to the following equations. 
\begin{equation}
{\mib w}^R_{\mib k}(0)={\mib u}^R_{\mib k}(0)
+\int_{k'}[{\rm coth}\frac{\epsilon'}{2T}
{\rm Im}V^R_{\mib k-\mib k'}(-\epsilon')
\frac{\partial {\rm Re}G^R_{\mib k'}(\epsilon')}{\partial \epsilon'}
+{\rm tanh}\frac{\epsilon'}{2T}
\frac{\partial {\rm Re}V^R_{\mib k-\mib k'}(-\epsilon')}{\partial \epsilon'}
{\rm Im}G^R_{\mib k'}(\epsilon')
]z_{\mib k'}{\mib w}^R_{\mib k'}(0)
\end{equation}
and 
\begin{equation}
{\mib u}^R_{\mib k}(0)=
\int_{k'}[{\rm coth}\frac{\epsilon'}{2T}
{\rm Im}V^R_{\mib k-\mib k'}(-\epsilon')
\frac{\partial {\rm Re}G^R_{\mib k'}(\epsilon')}{\partial \epsilon'}
+{\rm tanh}\frac{\epsilon'}{2T}
\frac{\partial {\rm Re}V^R_{\mib k-\mib k'}(-\epsilon')}{\partial \epsilon'}
{\rm Im}G^R_{\mib k'}(\epsilon')
]({\mib v}_{\mib k'}-{\mib v}_{\mib k}). 
\end{equation}
Here in the former equation we put $w^R_{\mib k'}(\epsilon')
/(1-\partial \Sigma^R_{\mib k'}(\epsilon')/\partial \epsilon')$ as 
$z_{\mib k'}w^R_{\mib k'}(0)$, because 
only the region of small $\epsilon'$ contributes 
to the integral owing to the factor 
$\partial G(\epsilon')/\partial \epsilon'$. 
The consistency between the numerical calculation and 
the analytical discussion in this section also verifies this replacement. 

As often taken in the spin fluctuation theories,~\cite{rf:24} 
we use the following approximation to $V^R_{\mib q}(\omega)$, 
\begin{equation}
V^R_{\mib q}(\omega)\simeq
\frac{\chi(\mib Q)}{1+\xi^2(\mib q-\mib Q)^2-{\rm i}\omega/\omega_{sf}}. 
\end{equation}
($\xi$ and $\omega_{sf}$ are the correlation length and 
the characteristic frequency of the antiferromagnetic 
spin fluctuation, respectively.) 
The following form of $G^R_{\mib k}(\epsilon)$ is also used, 
\begin{equation}
G^R_{\mib k}(\epsilon)\simeq
\frac{z_{\mib k}}{\epsilon-\xi^*_{\mib k}+{\rm i}\gamma_{\mib k}}. 
\end{equation}
Then the integration in eq.(4.7) is carried out at $T \rightarrow 0$ and 
it can be shown that the term which includes ${\rm Im}V^R$ vanishes 
and only the term which includes ${\rm Im}G^R$ remains. 
The result is the following equation, 
\begin{equation}
{\mib u}^R_{\mib k}(0)=
\frac{\pi \chi(\mib Q)}{2(2\pi)^2 v_{\rm F}\xi}
\int_{\rm FS}{\rm d}k'
\frac{1}{\pi\xi}\frac{\mib v_{\mib k'}-\mib v_{\mib k}}
{1/\xi^2+(\mib k-\mib k'-\mib Q)^2}. 
\end{equation}
Similarly we obtain 
\begin{equation}
{\mib w}^R_{\mib k}(0)={\mib u}^R_{\mib k}(0)+
\frac{\pi z_{\mib k}\chi(\mib Q)}{2(2\pi)^2 v_{\rm F}\xi}
\int_{\rm FS}{\rm d}k'
\frac{1}{\pi\xi}\frac{\mib w^R_{\mib k'}(0)}
{1/\xi^2+(\mib k-\mib k'-\mib Q)^2}. 
\end{equation}
($v_{\rm F}$ is the velocity at the Fermi surface.) 
Then in the limit of $\xi \rightarrow \infty$ 
the integral in these equations reduces to the delta function 
$\delta(\mib k-\mib k'-\mib Q)$. 
However, as $\xi$ is not infinitely large in real systems, 
we introduce the reducing factor. 
Then the following equations are obtained.
\begin{equation}
{\mib w}^R_{\mib k}(0)={\mib u}^R_{\mib k}(0)
+z_{\mib k}c\alpha{\mib w}^R_{\mib k-\mib Q}(0), 
\end{equation}
and 
\begin{equation}
{\mib u}^R_{\mib k}(0)=
c(\alpha{\mib v}_{\mib k-\mib Q}-{\mib v}_{\mib k}). 
\end{equation}
Here
$c:=\frac{\pi\chi(\mib Q)}{2(2\pi)^2v_{\rm F}\xi}$ 
and $\alpha$ is some constant which is the reducing factor 
owing to the momentum dependence of $V^R_{\mib q}(0)$ and 
$0<\alpha<1$. 
By using $(v_{\mib k-\mib Q x},v_{\mib k-\mib Q y})
=-(v_{\mib k y},v_{\mib k x})$, the above equations are 
solved as, 
\begin{equation}
w^R_{\mib k x}(0)=
\frac{u^R_{\mib k x}(0)-zc\alpha u^R_{\mib k y}(0)}
{1-(zc\alpha)^2}
\end{equation}
and 
\begin{equation}
u^R_{\mib k x}(0)=-c(v_{\mib k x}+\alpha v_{\mib k y}). 
\end{equation}
The equation for $y$-component is written by 
exchanging $x$ with $y$ in the above equations. 
To see the behavior of the current on the Fermi surface 
we analyze these equations on the typical three points, 
at two hot spots ($hs1$ and $hs2$) and at a cold spot ($cs$). 
The locations of these points are shown in Fig.3 of \S5. 
\begin{eqnarray}
u^R_{\mib k x}(0)\simeq -cv_{\mib k x}, 
u^R_{\mib k y}(0)\simeq -c\alpha v_{\mib k x} \;\;\; (at\; hs1),  \\
u^R_{\mib k x}(0)\simeq u^R_{\mib k y}(0)\simeq 
-c(1+\alpha)v_{\mib k x} \;\;\; (at\; cs), \\
u^R_{\mib k x}\simeq -c\alpha v_{\mib k y}, 
u^R_{\mib k y}\simeq -cv_{\mib k y} \;\;\; (at\; hs2). 
\end{eqnarray}
By using the above equations and 
$z\simeq (1+c)^{-1}$, $c>>1$ we obtain 
\begin{eqnarray}
j^*_{\mib k x}\simeq z_{\mib k}v_{\mib k x} \;\;\; (at\; hs1),\\
j^*_{\mib k x}\simeq z_{\mib k}v_{\mib k x} \;\;\; (at\; cs),\\
j^*_{\mib k x}\simeq \frac{-\alpha z_{\mib k}v_{\mib k y}}{1-\alpha^2}
\;\;\; (at\; hs2). 
\end{eqnarray}
From the above equations we can see that 
the growth of the antiferromagnetic spin fluctuation 
described by $\alpha$ affects mainly near $hs2$. 
Then we can see that the reduction of $1/\lambda^2$ 
accompanied by the spin fluctuation is mainly caused by $hs2$. 
It can be said that this point is strongly 
affected by the Umklapp scattering as can be seen in fig.2. 

On the other hand the temperature dependence of 
$1/\lambda^2$ at the low temperature is dominated 
by the point $cs$ and ${\rm d}\lambda^{-2}/{\rm d}T|_{T=0}$ is determined 
by $j^{*2}$ unlike $j^*$ in the case of $\lambda^{-2}$ at absolute zero. 
These properties are shown in the formula of 
${\rm d}\lambda^{-2}/{\rm d}T$ at absolute zero 
derived from eqs.(2.2) and (2.4) as 
\begin{equation} 
\frac{\rm d}{{\rm d} T}\left(\frac{1}{\lambda^2_{\mu\nu}(T)}\right)_{T=0}
\propto \int_{\rm FS}\frac{{\rm d} S_k}{2\pi^2|{\mib v}^*(\mib k)|}
j^*_{\mu}(\mib k)\left(-\frac{{\rm d} Y(\mib k;T)}{{\rm d} T}\right)_{T=0}
j^*_{\nu}(\mib k). 
\end{equation}
In this equation the main contribution comes from 
near point $cs$ because of the pairing 
symmetry of high-$T_{\rm c}$ cuprates. 
Then this equation indicates that the reduction 
of $1/\lambda^2(0)$ accompanied by the spin fluctuation 
does not mean that the 
reduction of $({\rm d}/{\rm d}T)\lambda^{-2}(0)$ because 
$j^*$ at $cs$ is not affected by the growth of the spin fluctuation 
as can be seen from eq.(4.20). 
This is also confirmed numerically in \S5.3. 

\subsection{Nature of Umklapp process}
The large value of the Umklapp term in the case of 
strong antiferromagnetic spin fluctuation 
is understood by considering the extreme example as 
follows. 
If $V_{\mib q}(\omega)$ and $W_{\mib q}(\omega)$ 
are, respectively, replaced by $V\delta(\mib q-\mib Q)\delta(\omega)$ and 
$W\delta(\mib q-\mib Q)\delta(\omega)$, then 
${\mib u}_{\mib k}(\epsilon)$ is written as 
\begin{eqnarray}
{\mib u}_{\mib k}(\epsilon)&=&
V\int_{k'}G_{\mib k-\mib Q}(\epsilon)G_{\mib k'-\mib Q}(\epsilon')
\frac{\partial G_{\mib k'}(\epsilon')}{\partial \epsilon'}
(\mib v_{\mib k}-\mib v_{\mib k-\mib Q}) \nonumber\\
&=&2V\int_{k'}G_{\mib k-\mib Q}(\epsilon)G_{\mib k'-\mib Q}(\epsilon')
\frac{\partial G_{\mib k'}(\epsilon')}{\partial \epsilon'}
\mib v_{\mib k}. 
\end{eqnarray}
Here $\mib v_{\mib k-\mib Q}$ is replaced by $-\mib v_{\mib k}$. 

As the $\mib q$-dependence of $V_{\mib q}(\omega)$ and 
$W_{\mib q}(\omega)$ is weakened, it is easily seen that 
the magnitude of the above $\mib u_{\mib k}(\epsilon)$ becomes 
small. 
On the other hand if $V_{\mib q}(\omega)$ and $W_{\mib q}(\omega)$ 
have sharp peaks at $\mib q=\mib 0$ $\omega=0$, i.e. 
in the case of strong ferromagnetic fluctuation, 
(though this is not the case of high-$T_{\rm c}$, we consider 
it as just an example to illustrate the effect of the Umklapp scattering) 
it can be seen that the Umklapp term vanishes 
as below. 
\begin{eqnarray}
{\mib u}_{\mib k}(\epsilon)&=&
V\int_{k'}G_{\mib k}(\epsilon)G_{\mib k'}(\epsilon')
\frac{\partial G_{\mib k'}(\epsilon')}{\partial \epsilon'}
(\mib v_{\mib k}-\mib v_{\mib k}) \nonumber\\
&=&0. 
\end{eqnarray}

\begin{figure}
\epsfile{file=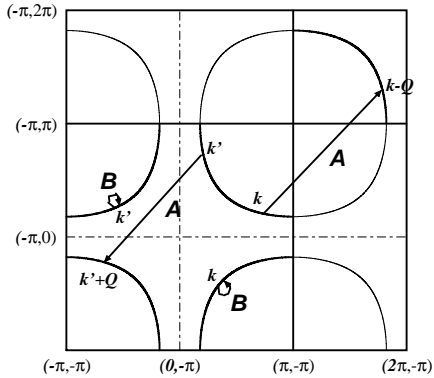,height=6.0cm}
\caption{The characteristic scattering processes with and 
without the Umklapp scattering}
\label{fig:2}
\end{figure}

These behaviors can be understood intuitively by 
drawing the processes like in Fig.~\ref{fig:2}. 
The scattering process is $(k,k')\rightarrow(k-q,k'+q)$, and then 
by using the notation in this figure the case of the strong 
antiferromagnetic spin fluctuation corresponds to the case where 
the process of $A$ dominates and the case of the strong 
ferromagnetic spin fluctuation corresponds to the case where 
the process of $B$ is dominant. 
The process $A$ is clearly the Umklapp process and the process 
$B$ means that the electron goes back to the beginning point on the 
Fermi surface and this process has no contribution to 
${\mib u}_{\mib k}(\epsilon)$. 
Both processes are equally contained in the case where 
the momentum dependence of the irreducible four point vertex is weak. 
In the actual system both processes exists and which type of 
these processes is dominant is the factor for determining 
whether the current $j^*<v^*$ or not. 

\subsection{Backflow}

In \S 2 it is shown that the backflow can be estimated 
from the two kinds of the derivative of self-energy by $\mib k$, 
$\partial \Sigma_{\mib k}^R(0)/\partial \mib k$ 
and ${\rm d} \Sigma_{\mib k}^R(0)/{\rm d} \mib k$. 
The integral equations which are obeyed by these quantity is 
written as follows; 
\begin{eqnarray}
\frac{\partial \Sigma^R_{\mib k}(0)}{\partial \mib k}
&=&\int_{k'}[{\rm coth}\frac{\epsilon'}{2T}
{\rm Im}V^R_{\mib k-\mib k'}(-\epsilon')
\frac{\partial {\rm Re}G^R_{\mib k'}(\epsilon')}{\partial \epsilon'}
+{\rm tanh}\frac{\epsilon'}{2T}
\frac{\partial {\rm Re}V^R_{\mib k-\mib k'}(-\epsilon')}{\partial \epsilon'}
{\rm Im}G^R_{\mib k'}(\epsilon') \nonumber\\
&+&\frac{\partial}{\partial \epsilon'}
\left({\rm tanh}\frac{\epsilon'}{2T}\right)
{\rm Re}V^R_{\mib k-\mib k'}(-\epsilon')
{\rm Im}G^R_{\mib k'}(\epsilon')
]z_{\mib k'}\left({\mib v}_{\mib k'}
+\frac{\partial \Sigma^R_{\mib k'}(0)}{\partial k'}
\right)
\end{eqnarray}
and 
\begin{eqnarray}
\frac{{\rm d}\Sigma^R_{\mib k}(0)}{{\rm d} \mib k}
&=&\int_{k'}[{\rm coth}\frac{\epsilon'}{2T}
{\rm Im}V^R_{\mib k-\mib k'}(-\epsilon')
\frac{\partial {\rm Re}G^R_{\mib k'}(\epsilon')}{\partial \epsilon'}
+{\rm tanh}\frac{\epsilon'}{2T}
\frac{\partial {\rm Re}V^R_{\mib k-\mib k'}(-\epsilon')}{\partial \epsilon'}
{\rm Im}G^R_{\mib k'}(\epsilon')]
z_{\mib k'}\left({\mib v}_{\mib k'}
+\frac{{\rm d}\Sigma^R_{\mib k'}(0)}{{\rm d} k'}
\right). \nonumber\\
&&
\end{eqnarray}
Here $\Sigma^R_{\mib k'}(\epsilon')$ is replaced by $\Sigma^R_{\mib k'}(0)$ 
by considering that small $\epsilon'$ region dominates in the
integral as before and terms including $W_{\mib q}(\omega)$ 
can be neglected for the same reason in the previous subsection. 
The main difference between the above two equation is the presence of 
the third term in eq.(4.25) and this is the difference between the 
$k$ and $\omega$-limits in the finite temperature formalism. 
The integral of the first term and the second term in eqs.(4.25) 
and (4.26) are carried out in the same way in eq.(4.7) and 
it can be seen that the third integration in eq.(4.25) has opposite sign 
and twice in magnitude as compared with the second one. 
Then the eqs.(4.25) and (4.26) reduce to 
\begin{equation}
\frac{\partial \Sigma^R_{\mib k}(0)}{\partial \mib k}
=-cz_{\mib k}\alpha(\mib v_{\mib k-\mib Q}+
\frac{\partial \Sigma^R_{\mib k-\mib Q}(0)}{\partial \mib k})
\end{equation}
and 
\begin{equation}
\frac{{\rm d} \Sigma^R_{\mib k}(0)}{{\rm d} \mib k}
=cz_{\mib k}\alpha(\mib v_{\mib k-\mib Q}+
\frac{\partial \Sigma^R_{\mib k-\mib Q}(0)}{\partial \mib k}). 
\end{equation}
By applying the argument in \S4.2 to the above equations 
we obtain the following behavior of these. 
\begin{equation}
\frac{\partial \Sigma^R_{\mib k}(0)}{\partial k_x}
\simeq \frac{\alpha(v_{\mib k y}+\alpha v_{\mib k x})}{1-\alpha^2}, 
\end{equation}
\begin{equation}
\frac{{\rm d} \Sigma^R_{\mib k}(0)}{{\rm d}k_x}
\simeq \frac{\alpha(-v_{\mib k y}+\alpha v_{\mib k x})}{1-\alpha^2}.  
\end{equation}
Then the backflow defined in \S2 is derived as follows. 
\begin{equation}
\frac{B_{\mib k x}}{z_{\mib k}}\simeq 
-\frac{2\alpha v_{\mib k y}}{1-\alpha^2}. 
\end{equation}
From these equations it can be seen that the backflow is 
negative in the case of $v_{\mib k y} \geq 0$ 
and the magnitude of this becomes large as 
the spin fluctuation grows (i.e. $\alpha$ has the larger value). 
These considerations indicate that $j^*_{\mib k x}$ 
is smaller than $v^*_{\mib k x}$ in magnitude 
and $\lambda^{-2}\leq 4\pi e^2 n/m^*$ follows this. 

\section{Results of Numerical Calculation}

In this section we present the results of 
numerical calculations based on the 
model and approximation presented in 
the previous section. 

The first Brillouin zone is divided into 
$128\times 128$ mesh and the roughness 
in the following figures of the Fermi surface and 
the velocity is caused by this mesh structure. 
The hole doping level is taken as 
$\delta=0.15$ in the figure with no indication. 

\begin{figure}
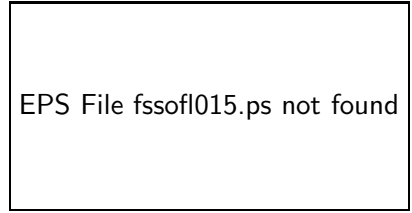

\epsfile{file=fssofl015,height=6.0cm}
\caption{Fermi surfaces with the hole doping 
$\delta=0.15$, calculated by using SC-SOPT and FLEX 
approximations.}
\label{fig:3}
\end{figure}

Firstly (a quarter of) the Fermi surfaces in the first 
Brillouin zone with the hole doping $\delta=0.15$, 
calculationed with SC-SOPT and FLEX approximations 
are shown in Fig.~\ref{fig:3}. 
In this figure $hs1$ and $hs2$ mean the hot spots 
which are the intersection points between the Fermi surface 
and the magnetic Brillouin zone (the line connecting 
$(\pi,0)$ with $(0,\pi)$ in this figure). 
These are the points connected by other points on Fermi surface 
by the vector ${\mib Q}=(\pi,\pi)$, and $cs$ means the cold spot 
which is the point (on the Fermi surface) 
most far from the magnetic Brillouin zone. 
These notations are the same as in some papers on 
high-$T_{\rm c}$ cuprates.~\cite{rf:9} 
It is also known that the Fermi surface is deformed 
by the correlation effect. 
From this figure it is noted that one with FLEX 
is deformed much to the magnetic Brillouin zone 
than the other with SC-SOPT. 
This is because the FLEX approximation includes 
the effect of the spin fluctuation effect much. 

\subsection{Momentum dependence of the current}

From here the only $x$-component of the vector is shown in the figure. 
The $y$-component at $(k_x,k_y)$ is identified with the 
$x$-component at $(k_y,k_x)$. 
$j^*_{\mib k x}$ in SC-SOPT is 
shown in Fig.4(a) with renormalized velocity 
$v^*_{\mib k x}$ and bare velocity $v_{\mib k x}$. 
\begin{figure}
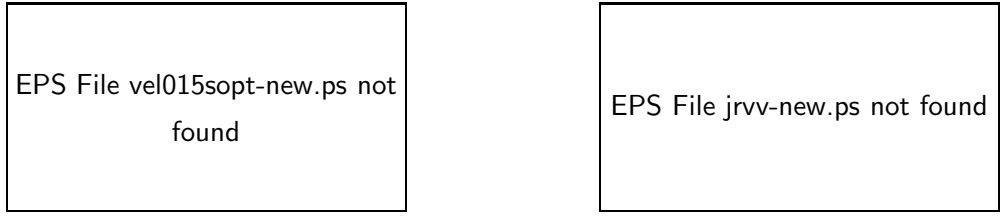

  \begin{minipage}[t]{.22\textwidth}
\epsfile{file=vel015sopt-new,height=6.0cm}
  \end{minipage}
  \hspace{4cm}
  \begin{minipage}[t]{.22\textwidth}
\epsfile{file=jrvv-new,height=6.0cm}
  \end{minipage}
\caption{The momentum dependence of $j^*_{\mib k x}$,$v^*_{\mib k x}$ and 
$v_{\mib k x}$ in (a) SC-SOPT and (b) FLEX}
\label{fig:4}
\end{figure}
From the argument given in \S4.1 
the current $j^*_{\mib k x}$ is not expected to 
be much different from $v^*_{\mib k x}\simeq 
z_{\mib k}v_{\mib k x}$ and this expectation is verified. 
The term $\partial \Sigma_{\mib k}(0)/\partial k_x$ in 
the renormalized velocity $v^*_{\mib k x}$ is negligible 
in the case of weak momentum dependence of the irreducible 
four point vertex. 
To put it more precisely, 
replacing $j^*_{\mib k x}$ by $v^*_{\mib k x}$, 
like in ref.3, is not verified exactly 
but only approximately with a condition given above on the irreducible 
four point vertex. 
$j^*_{\mib k x}$ in FLEX 
is shown in Fig.4(b). 
From this figure it can be seen that the current 
is different so much from the renormalized velocity, 
unlike in the case of SC-SOPT, owing to the 
strong spin fluctuation. 
There are two reasons of this difference. 
One is that $j^*_{\mib k x}$ is reduced by the large negative 
value of $w_{\mib k x}$ which originates from the large $u_{\mib k x}$. 
The other is that $v^*_{\mib k x}$ increases due to the 
positive value of $\partial \Sigma_{\mib k}(0)/\partial k_x$. 
The latter point manifests itself in $v^*_{\mib k x}>v_{\mib k x}$ at $hs2$. 
Both of these originate from the large spin fluctuation 
and in this case it is not allowed to approximate 
$j^*_{\mib k x}\simeq z_{\mib k}v_{\mib k x}$, as in SC-SOPT. 
The above mentioned difference between $j^*_{\mib k x}$ and 
$v^*_{\mib k x}$ also numerically verifies the discussion 
on the backflow in \S 4.4. 

\subsection{$\lambda^{-2}$}

Based on the above behaviors of $j^*_{\mib k x}$ 
the doping dependence of $\lambda^{-2}$ is presented. 
The result of SC-SOPT is shown in Fig.5(a). 
\begin{figure}
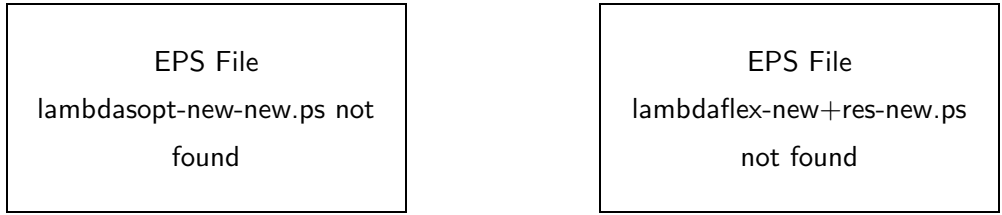

  \begin{minipage}[t]{.22\textwidth}
\epsfile{file=lambdasopt-new-new,height=6.0cm}
  \end{minipage}
  \hspace{4cm}
  \begin{minipage}[t]{.22\textwidth}
\epsfile{file=lambdaflex-new+res-new,height=6.0cm}
  \end{minipage}
\caption{The integrals over the Fermi surface, 
$\int_{\rm FS}\frac{{\rm d}k}{2\pi^2|\mib v^*_{\mib k}|}
v^*_{\mib k x}X$, in (a) SC-SOPT and (b) FLEX. 
Here $X=j^*_{\mib k x}$, $z_{\mib k}v_{\mib k x}$ and $v_{\mib k x}$.}
\label{fig:5}
\end{figure}
From this figure it can be seen that $\lambda^{-2}$, 
which is proportional to the case of $X=j^*_{\mib k x}$, 
is approximately described by $n/m^*$, 
which is proportional to the case of $X=z_{\mib k}v_{\mib k x}$. 
This is because the momentum dependence of the irreducible 
four point vertex is weak as discussed in \S3 and \S4.1. 
It is also noted that 
the value of the case $X=v_{\mib k x}$ decreases 
as the hole doping $\delta$ increases. 
This is because the case $X=v_{\mib k x}$ corresponds to 
$n/m$, $n$ means the effective carrier and increases 
owing to the increase of the effective carrier density 
$1-\delta$. ($\delta=0$ means the half-filled case 
and therefore the effective carrier density is 1.) 
The decrease of $n/m$ as $\delta$ increases 
indicates that $n/m^*=zn/m$ (the case of $X=v^*_{\mib k x}$) 
also decreases because the doping dependence of $z$ is 
rather weaker than the change of $n$. 
The above results are considered to explain the experimental 
doping dependence of $\lambda^{-2}$ in the overdoped 
region because the perturbation scheme is valid 
due to the weakness of the spin fluctuation in this region. 
The calculations by FLEX is shown in Fig.5(b). 
Unlike the case of SC-SOPT, it can be seen that the 
value and the doping dependence of the case $X=j^*_{\mib k x}$ 
is very different from those of the case $X=z_{\mib k}v_{\mib k x}$. 
This is because the spin fluctuation highly renormalizes the 
current. 
The doping dependence of the case $X=z_{\mib k}v_{\mib k x}$ 
is rather weak compared with the case of SC-SOPT owing to 
decreasing of the renormalization factor in the low doping. 
On the other hand the value of the case $X=j^*_{\mib k x}$ 
decreases as the $\delta$ decreases because the spin 
fluctuation grows in the low doping region. 
This behavior of $\lambda^{-2}$ (the case of $X=j^*_{\mib k x}$) 
is considered to explain the experimental doping dependence 
in the optimal and the underdoped regions. 

While the physical meaning and a justification of the usage 
of the perturbation theory on the overdoped region 
and the spin fluctuation theory on the optimal and 
the underdoped regions is discussed in \S7, 
we discuss the following two points. 
One of these is that by discussing the above results conversely, 
it can be said that the peak at the slightly overdoped 
region not at the optimal doping as experimentaly observed 
suggest that the spin flucutation begins to grow 
in this region, and therefore this is 
reflected in other quantities like the 
one-particle spectrum.~\cite{rf:11} 
The other is about the renormalization factor $z$. 
The calculated $z$ by SC-SOPT and FLEX are not 
smoothly connected. This suggests that the 
higher order terms in the perturbation expansion 
are needed to calculate the doping level itself of the 
peak in $\lambda^{-2}$ as discussed in \S7, 
while the explanation on the behaviors of $\lambda^{-2}$ 
on both sides of the peak is not modified. 

\subsection{On the slope of $\lambda^{-2}$ at absolute zero}

In high-$T_{\rm c}$ cuprates it is known that 
at the low temperature 
$\lambda^{-2}$ decreases linearly with $T$ 
because of the line nodes in the superconducting gap. 
If we assume the isotropic case, 
the value of $\lambda^{-2}$ at $T=0$ is given by, 
\begin{equation}
\lambda^{-2} \propto 
\frac{n}{m^*}\left(1+\frac{F_1^s}{2}\right), 
\end{equation}
which comes from the fact that $\lambda^{-2}$ is 
linear in $j^*$. 
On the other hand from eq.(4.22) 
it can be seen that the coefficient of $T$ in $\lambda^{-2}$ 
at the low temperature is given by 
\begin{equation}
\frac{{\rm d}\lambda^{-2}}{{\rm d}T}
\propto
\frac{n}{m^*}\left(1+\frac{F_1^s}{2}\right)^2, 
\end{equation}
which comes from the fact that 
${\rm d}\lambda^{-2}/{\rm d}T$ is square in $j^*$ 
in this case. 
Then it is expected that if the decreasing of $\lambda^{-2}$ 
is attributed to small $1+F_1^2/2$ as $\delta$ decreases, 
then the rate of decreasing of ${\rm d}\lambda^{-2}/{\rm d}T$ 
is expected to be more rapid than $\lambda^{-2}$ owing to $(1+F_1^2/2)^2$. 
This expectation is denied by the experimental results 
({\it e.g.} refs.25 and 26). 
Here we clarify that this failure in the explanation 
for the experiments is not caused by the failure of the 
explanation based on the Fermi liquid theory 
but owning to the use of the isotropic model 
which is unrealistic as understood from the previous sections. 

The slope ${\rm d}\lambda^{-2}/{\rm d}T$ at $T=0$ 
comes from the derivative of Yosida function by the 
temperature ${\rm d}Y(\mib k;T)/{\rm d}T$. 
From the fact that the excitations 
of the quasiparticles at the low temperature 
are mainly produced at the line nodes (which are equal to 
the cold spots in our system), 
${\rm d}Y(\mib k;T)/{\rm d}T$ has a large value 
at these points. 
Therefore it is allowed to consider the function $j^{*2}_{\mib k}$ 
only at the cold spots in ${\rm d}\lambda^{-2}/{\rm d}T|_{T=0}$. 
On the other hand for the value of $\lambda^{-2}$ at $T=0$ 
the whole value of $v^*_{\mib k x}j^*_{\mib k x}$ on the Fermi surface 
should be considered in practice. 

With the above consideration in mind 
we present the momentum dependence of $j^*_{\mib k x}$ 
at hole doping $\delta=0.10$ and $\delta=0.20$ in Fig.6(a). 
\begin{figure}
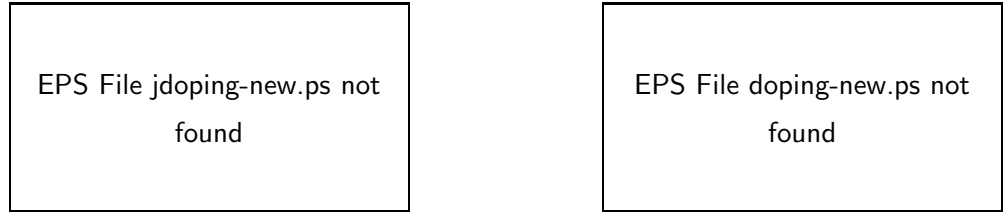

  \begin{minipage}[t]{.22\textwidth}
\epsfile{file=jdoping-new,height=6.0cm}
  \end{minipage}
  \hspace{4cm}
  \begin{minipage}[t]{.22\textwidth}
\epsfile{file=doping-new,height=6.0cm}
  \end{minipage}
\caption{(a) The momentum dependence of $j^*_{\mib k x}$ 
at $\delta=0.10$ and $\delta=0.20$ in FLEX. 
(b) The doping dependence of $j^*_{\mib k x}$ 
at the characteristic points $hs1$, $cs$ and $hs2$ 
on Fermi surface in FLEX.}
\label{fig:6}
\end{figure}
From this figure it can be seen that the doping 
dependence of the current $j^*_{\mib k x}$ around 
$\mib k=hs2$ is large while those around 
$\mib k=hs1,cs$ is weak. 
The doping dependences at these points 
are shown in Fig.6(b). 
From these figures it can be seen that the 
decrease of $\lambda^{-2}$ at $T=0$ presented in fig.5(b) 
is caused by the decrease of $j^*_{\mib k x}$ around the points 
$hs2$ while $j^*_{\mib k x}$ around the points 
$hs1$ and $cs$ doesn't contribute so much on the doping dependence 
of this quantity. 
On the other hand the small doping dependence of 
$j^*_{\mib k x}$ around $cs$ is reflected in the small 
doping dependence of ${\rm d}\lambda^{-2}/{\rm d}T$. 
These behaviors of $j^*_{\mib k x}$, with the anisotropy 
included, are considered to explain the experimental 
results of $\lambda^{-2}$ and 
${\rm d}\lambda^{-2}/{\rm d}T$ consistently. 

\section{Relation between Magnetic Field Penetration Depth and 
Superconducting Fluctuation}

The high-$T_{\rm c}$ cuprates is known to show the 
pseudogap phenomena in the underdoped region. 
Some authors are tried to explain this 
phenomenon by relating this with 
large value of the magnetic field penetration depth.~\cite{rf:4} 
The point of their argument is 
that by writing the phase only model 
the phase stiffness is proportional to the inverse 
of squared magnetic field penetration depth and 
then the large value of $\lambda$ {\it at absolute zero} 
in the underdoped region 
means that the phase fluctuation is dominant in  this region. 
In this section we investigate whether this assertion is correct. 

The GL model is written as 
\begin{equation}
{\cal H}=\int{\rm d}{\mib r}\left[
a(\epsilon_0|\psi|^2+\xi_0^2
|(-{\rm i}{\mib \nabla}+\frac{2\pi}{\Phi_0}{\mib A})\psi|^2)
+\frac{b}{2}|\psi|^4\right]. 
\end{equation}
Here 
$a:=\rho^*(0)$ ($\rho^*(0)$ is the density of states at Fermi level 
enhanced by $1/z$), $\epsilon_0:=(T-T_{\rm c})/T_{\rm c}$, 
$\xi_0:=\frac{7\zeta(3)}{48}\left(\frac{v_{\rm F}}{\pi T_{\rm
c}}\right)^2$, $\Phi_0$ is the flux quantum 
and $b:=\frac{7\zeta(3)}{8}\frac{\rho^*(0)}
{(\pi T_{\rm c})^2}$ by the microscopic calculation with an usual model 
where the attractive force is operated between renormalized 
quasiparticles. 
By using $\psi=|\psi|{\rm e}^{\rm i \phi}$ 
this model is reduced to a phase only model, 
\begin{equation}
{\cal H}=\int{\rm d}{\mib r}\left[
-\frac{a|\epsilon_0|^2}{2b}
+\frac{1}{8\pi}\lambda^{-2}(T)
\left({\mib A}+\frac{\Phi_0}{2\pi}{\mib \nabla}\phi
\right)^2\right]. 
\end{equation}
Here we put 
\begin{equation}
\lambda^{-2}(T)=\left(\frac{2\pi\xi_0}{\Phi_0}
\right)^2\frac{8\pi a^2|\epsilon_0|}{b}.  
\end{equation}
From this expression with the above parameter inserted 
it is seen that the $\lambda^{-2}$ is proportional to $z$. 
This is in contrast to $\lambda^{-2} \propto z(1+F_1^s/2)$ 
at absolute zero discussed in the previous sections. 
On the other hand, from eqs.(2.2) and (2.5) the magnetic field 
penetration depth near the transition temperature is given by
\begin{equation}
\frac{1}{\lambda^2_{\mu\nu}(T\simeq T_{\rm c})}
\propto
\int_{\rm FS}\frac{{\rm d} S_k}{2\pi^2|{\mib v}^*(\mib k)|}
v^*_{\mu}(\mib k)(1-Y(\mib k;T))v^*_{\nu}(\mib k). 
\end{equation}
$\epsilon_0$ term is derived from $1-Y(\mib k;T)$ and 
we neglected the second term in the right hand side of eq.(2.5) 
due to the smallness of $1-Y(\mib k;T)$ near $T_{\rm c}$. 
From this equation $\lambda^{-2}(T\simeq T_{\rm c})$ is 
proportional to the renormalization factor $z$ and 
this is consistent with the eq.(6.3) not with the 
expression of $\lambda^{-2}$ at $T=0$. 
This suggests that 
$F_1^s$ term which is one of the particle-hole 
four point vertices is not effective near $T_{\rm c}$ 
where the particle-particle four point vertex is remarkable, 
although this notion is much simplified and 
should be elaborate for the exact description. 

The above consideration indicates that the 
the smallness of the coefficient in the phase only model 
cannot be identified with the smallness of $\lambda^{-2}$ 
at the absolute zero even if the temperature dependence 
$\epsilon_0$ is excluded. 
This is originated from the fact that there is a non trivial 
temperature dependence in the vertex correction in the 
superconducting state. 
Therefore the small $\lambda^{-2}$ 
at absolute zero doesn't necessarily indicate that the 
thermal fluctuation around $T_{\rm c}$ is large 
because the renormalization factor in the phase 
model is not $z(1+F^s_1/2)$ but $z$. 
Then the the small $\lambda^{-2}$ cannot be 
directly related with large thermal fluctuation 
in the underdoped region although it is considered 
that the thermal fluctuation is large as experiments indicate, 
but the origin of the large fluctuation is composed of the 
factors like the strong coupling effect and 
the quasi two dimensionality.~\cite{rf:7} 

\section{Summary and Discussion}
In this paper we derived the general expression 
for the effect of the Umklapp scattering on the current 
and discussed the magnetic field penetration 
depth, particularly of high-$T_{\rm c}$ cuprates, 
both analytically and numerically. 

In \S2 a formula for the current carried by 
quasiparticles which explicitly expresses the 
effect of the Umklapp scattering is derived. 
This is basically derived from the 
usual expression for the current written by 
the renormalized velocity with the backflow. 
However to see the effect of the Umklapp scattering 
the above two quantities are not basic but 
this effect is written by the integrals over the 
whole energy scale not only over the Fermi surfaces. 
This is the basic difference between the 
current in the hydrodynamic region and in the collisionless region, 
and indicates that the backflow term 
is not necessarily the basic quantity in the lattice system. 
The advantages of our expression for the Umklapp term over the notion 
of the backflow are the followings. 
We can show that the upper limit 
for the current is the bare velocity and doesn't exceed this value 
and that the correlation effect is unified in the Umklapp term 
while by using the backflow notion the renormalized velocity 
also includes the correlation effect. 

In \S4 the Umklapp term, mainly of the high-$T_{\rm c}$ 
cuprates, is analytically discussed. 
In the highly overdoped region of the high-$T_{\rm c}$ cuprates 
the momentum dependence of the irreducible 
four point vertex is weak though it exists, 
and by neglecting this small term the Umklapp term 
is written only by the self-energy term. 
Therefore the current is expressed only by the 
product of the renormalization factor and the bare 
velocity and then the $\lambda^{-2}$ is proportional 
to $n/m^*$. 
By this fact the reason why a misunderstanding 
of $\lambda^{-2}\propto n/m^*$ like ref.3 has not been 
confronted with a difficulty so far is understood. 
In the optimal and underdoped regions 
it is known that there exists the strong spin fluctuation, 
and the temperature-linear resistivity and 
the temperature dependent Hall coefficient can be 
derived by using this feature. 
Therefore it is reasonable to start with the spin fluctuation 
model to discuss the magnetic field penetration depth too. 
It is found that the strong momentum 
dependence of the irreducible four point vertex induces 
the large vertex correction in addition to the self-energy 
term and this correction makes the Umklapp term 
dependent on the points on the Fermi surface. 
Therefore at some points the current is not written by 
the product of the renormalization factor and the bare 
velocity unlike the case of the highly overdoped region, 
but reflects the strong Umklapp scattering at these points. 
These considerations indicate that the strong antiferromagnetic 
spin fluctuation can be identified with the strong Umklapp scattering 
as the effect on the current. 
The other characteristic results 
are that the backflow always takes an opposite sign to 
the bare velocity and the partial derivative of the 
self-energy by the momentum takes the same sign as the bare velocity 
with non-negligible magnitude. 
The latter point indicates that the renormalized velocity 
is not so simple and supports the above assertion 
that the dividing the current to the renormalized velocity 
and the backflow is not useful. 

The main results obtained by numerical calculations in \S5 
are that the doping dependence of $\lambda^{-2}$ in experiments 
can be explained by using the perturbation scheme in the overdoped 
region and the spin fluctuation theory in the optimal and 
underdoped regions, and that the relation between $\lambda^{-2}$ and 
${\rm d}\lambda^{-2}/{\rm d}T$ at absolute zero indicated 
by experiments, which was failed to explain by the isotropic 
model,~\cite{rf:13} is explained by using the anisotropic model based 
on the spin fluctuation theory. 
As regards with the latter point the way to recover 
the consistency between these two quantities based 
on the isotropic model is indicated in ref.25 
by using the angle resolved photo-emission spectroscopy 
(ARPES) experiments. 
It is summarized as that ARPES experiments indicate that 
the form of the gap function deforms from the $d_{x^2-y^2}$-pairing 
with underdoping and this suggests that the magnitude of the gap 
around the cold spot decreases and then the excitations 
around this point increases with underdoping. 
Although this behavior may be possible in the underdoped 
region, the calculated gap with the FLEX approximation 
suggests that the deformation from the $d_{x^2-y^2}$-pairing 
is not so drastic as this experiments suggests. 
While one of the reasons for this is the inaccuracy of the 
approximation, the other is that the ARPES experiments at the cold 
spot where the dispersion is sharp, 
cannot be accurate to discuss the form of the gap function 
around the node because of the limited resolution in ARPES. 
In spite of these facts the anisotropy of the system is not obviously 
negligible and it is needed for considering the qualitative estimation 
for ${\rm d}\lambda^{-2}/{\rm d}T$. 

The explanation on the doping dependence of $\lambda^{-2}$ 
needs some notions. It is desirable not to rely 
on the specific approximations, but this cannot be done. 
Then it is needed to consider what is physically the most 
appropriate method. 
Because we stand upon the Fermi liquid theory, 
the perturbation scheme should be taken at first. 
In high-$T_{\rm c}$ cuprates the on-site Coulomb interaction 
in the Hubbard model is large, and then it may be considered 
that all higher order terms is effective in the expansion. 
The reason why the some low order terms is sufficient in 
the overdoped region and the spin fluctuation theory 
is appropriate (only approximately) in the optimal and 
the underdoped region, is explained as follows. 
In the study on the single impurity system, it is shown 
that in the perturbation expansion by the on-site Coulomb 
interaction $U$, the coefficient of $U^n$ ($n$ is the order 
of the expansion) becomes smaller as $n$ becomes large.~\cite{rf:27} 
This indicate that it is practically sufficient to 
take only the low order terms in $U$, 
even if $U$ is somewhat large, in contrast to 
the mean field treatment. 
The notable point is that there is no momentum dependence 
in the single impurity system. 
This indicate that it is reasonable to take 
the low order perturbation scheme as an approximation 
in the overdoped region where the momentum dependence 
of the vertex is rather weak owing to its filling. 
With underdoping the momentum dependence of the vertex 
becomes strong because the system approaches to 
the antiferromagnetic phase. 
Therefore in this case taking only the low order 
terms is not appropriate owing to the fact that 
the cancellation between higher order terms is not effective, 
but the specific mode is considered to be effective up 
to higher order terms and this leads to the antiferromagnetic 
spin fluctuation. 
This is the reason why we take the spin fluctuation 
theory as an approximation in the optimal and underdoped regions. 
It is also the reason why FLEX approximation 
is inappropriate in the overdoped region because 
other higher order terms which should cancel 
the RPA type terms in the case of the weak momentum dependence 
are not included. 
The above qualitative discussions should be verified 
by explicitly calculating the higher order terms with various doping levels. 
This is an important feature problem. 

The other notions with the FLEX approximation is that 
this approximation is not good to describe the Mott 
transition.~\cite{rf:28} 
The main fault of this approximation is lack 
of the Hubbard peaks in the density of states.~\cite{rf:29} 
However it is the momentum dependence at the low energy 
that is important for our explanation on the current and 
the magnetic field penetration depth, and therefore 
main results in this paper is not changed qualitatively 
by the existence of the high-energy structure like 
the Hubbard peak while the improvement of the density 
of states would slightly affect the quantitative results. 
  
\section*{Acknowledgment}
Numerical computation in this work was carried out at the 
Yukawa Institute Computer Facility.

\appendix
\section{Expressions at Finite Temperature}
At finite temperature the corresponding equation of eq.(2.20) is 
(${\mib w}^R_{\mib k}(\epsilon):=
\frac{{\rm d}\Sigma^R_{\mib k}(\epsilon)}{{\rm d}\mib k}+
\frac{\partial\Sigma^R_{\mib k}(\epsilon)}{\partial\epsilon}
{\mib v}_{\mib k}$), 
\begin{eqnarray}
{\mib w}^R_{\mib k}(\epsilon)
&=&\frac{1}{2{\rm i}}\int_{k'}\left[
\Im^{(11)}_{\mib k,\mib k'}(\epsilon,\epsilon')
\left(-\frac{\partial G^R_{\mib k'}(\epsilon')}{\partial \epsilon'}\right)
+\Im^{(13)}_{\mib k,\mib k'}(\epsilon,\epsilon')
\left(-\frac{\partial G^A_{\mib k'}(\epsilon')}{\partial \epsilon'}\right)
\right]
({\mib v}_{\mib k'}-{\mib v}_{\mib k}) \nonumber\\
&-&\int_{k'}\frac{1}{2T}\frac{1}{{\rm cosh}^2\epsilon'/2T}
\Gamma^{(12)}_{\mib k,\mib k'}(\epsilon,\epsilon')
{\rm Im}G^R_{\mib k'}(\epsilon')({\mib v}_{\mib k'}-{\mib v}_{\mib k})
\nonumber\\
&+&\frac{1}{2{\rm i}}\int_{k'}
\left[
\Im^{(11)}_{\mib k,\mib k'}(\epsilon,\epsilon')
\left(-\frac{\partial G^R_{\mib k'}(\epsilon')}{\partial \epsilon'}\right)
\frac{{\mib w}^R_{\mib k'}(\epsilon')}
{1-\partial \Sigma^R_{\mib k'}(\epsilon')/\partial \epsilon'}
+\Im^{(13)}_{\mib k,\mib k'}(\epsilon,\epsilon')
\left(-\frac{\partial G^A_{\mib k'}(\epsilon')}{\partial \epsilon'}\right)
\frac{{\mib w}^A_{\mib k'}(\epsilon')}
{1-\partial \Sigma^A_{\mib k'}(\epsilon')/\partial \epsilon'}
\right] \nonumber\\
&-&\int_{k'}\frac{1}{2T}\frac{1}{{\rm cosh}^2\epsilon'/2T}
\Gamma^{(12)}_{\mib k,\mib k'}(\epsilon,\epsilon')
{\rm Im}G^R_{\mib k'}(\epsilon')
\frac{{\mib w}^R_{\mib k'}(\epsilon')}
{1-\partial \Sigma^R_{\mib k'}(\epsilon')/\partial \epsilon'}. 
\end{eqnarray}
Here the meanings of $(11)$, etc. of $\Im$ are 
same as in \'Eliashberg's paper~\cite{rf:30} and these are 
\begin{equation}
\Im^{(11)}_{\mib k,\mib k'}(\epsilon,\epsilon')=
{\rm tanh}\frac{\epsilon'}{2T}
[\Gamma^{(a)}_{\mib k,\mib k'}(\epsilon,\epsilon')
+\Gamma^{(b)}_{\mib k,\mib k'}(\epsilon,\epsilon')]
+2{\rm i}{\rm cosh}\frac{\epsilon-\epsilon'}{2T}
\Delta^{(a)}_{\mib k,\mib k'}(\epsilon,\epsilon'), 
\end{equation}
\begin{eqnarray}
\frac{\partial \Im^{(12)}_{\mib k,\mib k'}(\epsilon,\epsilon';\omega)}
{\partial \omega}|_{\omega=0}
&=&\frac{1}{2T}\frac{1}{{\rm cosh}^2\epsilon'/2T}
\Gamma^{(12)}_{\mib k,\mib k'}(\epsilon,\epsilon') \\
&=&\frac{1}{2T}\frac{1}{{\rm cosh}^2\epsilon'/2T}
[\Gamma^{(a)}_{\mib k,\mib k'}(\epsilon,\epsilon')
+\Gamma^{(b)}_{\mib k,\mib k'}(\epsilon,\epsilon')]
\end{eqnarray}
and 
\begin{equation}
\Im^{(13)}_{\mib k,\mib k'}(\epsilon,\epsilon')=
-{\rm tanh}\frac{\epsilon'}{2T}
[\Gamma^{(a)}_{\mib k,\mib k'}(\epsilon,\epsilon')
+\Gamma^{(b)}_{\mib k,\mib k'}(\epsilon,\epsilon')]
+2{\rm i}{\rm cosh}\frac{\epsilon+\epsilon'}{2T}
\Delta^{(b)}_{\mib k,\mib k'}(\epsilon,\epsilon'). 
\end{equation}
Then
\begin{eqnarray}
{\mib w}^R_{\mib k}(\epsilon)
& = & {\mib u}^R_{\mib k}(\epsilon)+\int_{k'}
\{
{\rm tanh}\frac{\epsilon'}{2T}
\frac{\partial}{\partial \epsilon'}
[\Gamma^{(a)}_{\mib k,\mib k'}(\epsilon,\epsilon')
+\Gamma^{(b)}_{\mib k,\mib k'}(\epsilon,\epsilon')]
{\rm Im}G^R_{\mib k'}(\epsilon)
\frac{{\mib w}^R_{\mib k'}(\epsilon')}
{1-\partial \Sigma^R_{\mib k'}(\epsilon')/\partial \epsilon'}
\nonumber\\
& - & {\rm coth}\frac{\epsilon-\epsilon'}{2T}
\Delta^{(a)}_{\mib k,\mib k'}(\epsilon,\epsilon')
\frac{\partial G^R_{\mib k'}(\epsilon')}{\partial \epsilon'}
\frac{w^R_{\mib k'}(\epsilon')}
{1-\partial \Sigma^R_{\mib k'}(\epsilon')/\partial \epsilon'}
-{\rm coth}\frac{\epsilon+\epsilon'}{2T}
\Delta^{(b)}_{\mib k,\mib k'}(\epsilon,\epsilon')
\frac{\partial G^A_{\mib k'}(\epsilon')}{\partial \epsilon'}
\frac{{\mib w}^A_{\mib k'}(\epsilon')}
{1-\partial \Sigma^R_{\mib k'}(\epsilon')/\partial \epsilon'}. \nonumber\\
&& 
\end{eqnarray}
Here we put 
$\Gamma^{(a,b)}_{\mib k,\mib k'}(\epsilon,\epsilon')
={\rm Re}\Gamma^{(a,b)}_{\mib k,\mib k'}(\epsilon,\epsilon')
+{\rm i}\Delta^{(a,b)}_{\mib k,\mib k'}(\epsilon,\epsilon')$ as 
the four-point vertex which has a discontinuity along the cut 
${\rm Im}(\epsilon-\epsilon')=0$ or 
${\rm Im}(\epsilon+\epsilon')=0$ for the case of (a) or (b) in superscript, 
respectively. 
${\mib u}^R_{\mib k}(\epsilon)$ has the following form and 
it is seen that this quantity vanishes in the absence of 
Umklapp process. 
\begin{eqnarray}
{\mib u}^R_{\mib k}(\epsilon) & = & 
\int_{k'}\{{\rm coth}\frac{\epsilon-\epsilon'}{2T}
\Delta^{(a)}_{\mib k,\mib k'}(\epsilon,\epsilon')
\frac{\partial G^R_{\mib k'}(\epsilon')}{\partial \epsilon'}
+{\rm coth}\frac{\epsilon+\epsilon'}{2T}
\Delta^{(b)}_{\mib k,\mib k'}(\epsilon,\epsilon')
\frac{\partial G^A_{\mib k'}(\epsilon')}{\partial \epsilon'} \nonumber\\
&-&{\rm tanh}\frac{\epsilon'}{2T}
\frac{\partial}{\partial \epsilon'}
[\Gamma^{(a)}_{\mib k,\mib k'}(\epsilon,\epsilon')
+\Gamma^{(b)}_{\mib k,\mib k'}(\epsilon,\epsilon')]
{\rm Im}G^R_{\mib k'}(\epsilon')\}({\mib v}_{\mib k}-{\mib v}_{\mib k'}) \\
&=& -\int_{k'}\int_{q}\int\frac{{\rm d}x}{\pi}
\frac{{\rm cosh}\frac{x-\omega+\epsilon}{2T}}
{{\rm cosh}\frac{\omega-\epsilon}{2T}{\rm cosh}\frac{\epsilon'}{2T}
{\rm cosh}\frac{\epsilon'+x}{2T}}
|\Gamma_{\mib k,\mib k',\mib q}(\epsilon,\epsilon',\omega,x)|^2 \nonumber\\
&\times&\frac{{\rm Im}G^R_{k-q}(\epsilon-\omega)
{\rm Im}G^R_{k'}(\epsilon'){\rm Im}G^R_{k'+q}(\epsilon'+x)}
{(x-\omega-{\rm i}\delta)^2}({\mib v}_{\mib k-\mib q}+{\mib v}_{\mib k'+\mib q}
-{\mib v}_{\mib k'}-{\mib v}_{\mib k}). 
\end{eqnarray}
From eq.(A.7) to eq.(A.8) we used the spectral representation 
\begin{equation}
G^R_{\mib k}(\epsilon)=
\int\frac{{\rm d}x}{\pi}
\frac{{\rm Im}G^R_{\mib k}(x)}{x-\epsilon-{\rm i}\delta}
\end{equation}
(here $\delta$ is positive infinitesimal quantity) and 
the trivial relation between hyperbolic functions 
\begin{equation}
\left({\rm cosh}\frac{x}{2T}-{\rm tanh}\frac{\omega-\epsilon}{2T}\right)
\left({\rm tanh}\frac{\epsilon'}{2T}-{\rm tanh}\frac{x+\epsilon'}{2T}\right)
=-\frac{{\rm cosh}\frac{x-\omega+\epsilon}{2T}}
{{\rm cosh}\frac{\omega-\epsilon}{2T}{\rm cosh}\frac{\epsilon'}{2T}
{\rm cosh}\frac{\epsilon'+x}{2T}}
\end{equation}
From eq.(A.6) and eq.(A.8) it is obvious that in the continuum  
$w^R_{\mib k}(\epsilon)=0$ also holds at finite temperature 
as derived at absolute zero in \S2. 
By taking the SC-SOPT and FLEX approximations 
as example the validity of the above equation can be confirmed. 
In the latter approximation it is needed to use the exact relation 
${\rm Im}V^R_{\mib q}(\omega)=W_{\mib q}(\omega)
{\rm Im}\chi^R_{\mib q}(\omega)$.

\end{document}